\documentclass[
aps,
pra,
twocolumn,
superscriptaddress,
nofootinbib
]{revtex4-1}

\usepackage{pdfpages}
\usepackage{pgffor}
\makeatletter
\AtBeginDocument{\let\LS@rot\@undefined}
\makeatother

\usepackage{hyperref}
\usepackage{graphicx}
\usepackage{amsmath,braket}
\usepackage{dsfont}
\usepackage{upgreek}
\usepackage{qcircuit}

\DeclareMathOperator{\tr}{tr}

\newcommand{\smat}[4]{
    \scriptsize
    \begin{pmatrix}
    #1 & #2 \\
    #3 & #4
    \end{pmatrix}
}
\newcommand{\tab}[1]{\begin{tabular}{c} #1 \end{tabular}}
\newcommand{\id}{\mathds{1}}
\newcommand{\ketbra}[2]{
    \lvert #1 \rangle \! \langle #2 \rvert
}

\usepackage[utf8]{inputenc}
\usepackage[english]{babel}
\usepackage{hyperref}
\hypersetup{
    colorlinks=true,
    linkcolor=blue,
    filecolor=magenta,  
    citecolor=blue,
    urlcolor=blue,
}

\usepackage{listings} 
\usepackage[newfloat, frozencache]{minted}
\usepackage{subcaption}
\usepackage{ragged2e}
\DeclareCaptionJustification{justified}{\justifying}
\newenvironment{code}{\captionsetup{type=listing}}{}
\SetupFloatingEnvironment{listing}{name=Program}
\lstdefinestyle{codeblock}{
  basicstyle=\footnotesize,
  otherkeywords={with},
  breakatwhitespace=false,         
  breaklines=true,                 
  captionpos=b,                    
  keepspaces=true,
  language=Python,
  numbers=right,                    
  numbersep=5pt,                  
  showspaces=false,                
  showstringspaces=false,
  showtabs=false,                  
  tabsize=2,
  basicstyle=\ttfamily\footnotesize
}
\newmintedfile[pythonfile]{python}{
	bgcolor=gray!6,
	fontfamily=tt,
	gobble=0,
	fontsize=\footnotesize,
	breaklines=true,
	frame=lines,
	framesep=1mm,
	funcnamehighlighting=true,
	tabsize=2,
	obeytabs=false,
	mathescape=false
	samepage=false, 
	showspaces=false,
	showtabs =false,
	texcl=false,
	linenos=false,
	xleftmargin=1pt,
	numbersep=1pt
}

\usepackage{array,mathtools,amssymb,booktabs}
\DeclarePairedDelimiter{\abs}{\lvert}{\rvert}
\newcolumntype{C}{>{$}c<{$}}
\AtBeginDocument{
    \heavyrulewidth=.08em
    \lightrulewidth=.05em
    \cmidrulewidth=.03em
    \belowrulesep=.65ex
    \belowbottomsep=0pt
    \aboverulesep=.4ex
    \abovetopsep=0pt
    \cmidrulesep=\doublerulesep
    \cmidrulekern=.5em
    \defaultaddspace=.5em
}

\begin{document}

\title{A distributed simulation framework for quantum networks and channels}
\author{Ben Bartlett}
\email{benbartlett@stanford.edu}
\affiliation{Department of Applied Physics, Stanford University, Stanford, CA 94305, USA}
\affiliation{INQNET at AT\&T Foundry, 260 Homer Ave \#102, Palo Alto, CA 94301, USA}

\begin{abstract}
We introduce the \emph{Simulator for Quantum Networks and Channels} (\texttt{SQUANCH}), an open-source Python library for creating parallelized simulations of distributed quantum information processing. The framework includes many features of a general-purpose quantum computing simulator, but it is optimized specifically for simulating quantum networks. It includes functionality to allow users to easily design complex multi-party quantum networks, extensible classes for modeling noisy quantum channels, and a multiprocessed NumPy backend for performant simulations. We present an overview of the structure of the library, describing how the various API elements represent the underlying physics and providing simple usage examples for each module. Finally, we present several demonstrations of canonical quantum information protocols implemented using this framework.
\end{abstract}

\maketitle

\section{Introduction}

For decades, universal fault-tolerant quantum computing has promised to bring revolutionary changes to many important applications, offering significant computational speedups for problems such as integer factorization \cite{Shor1994PolynomialComputer}, functional inversion \cite{Grover1996ASearch}, finding expectations of solutions for linear systems \cite{Harrow2009QuantumEquations}, and estimating the ground state energy of complex molecules \cite{Lanyon2010TowardsComputer, Aspuru-Guzik2005Chemistry:Energies}. Although near-term quantum technology is limited in complexity by noise \cite{Preskill2018QuantumBeyond}, sub-universal quantum and hybrid computing devices have found application in a variety of fields ranging from traffic optimization \cite{Neukart2017TrafficAnnealer} to particle physics \cite{Mott2017SolvingLearning}. As quantum technologies advance, we may see the development of \emph{quantum networks}, which can distribute quantum information between remote nodes and promise cryptographically-secure communication \cite{Alleaume2014UsingSurvey}.

The rapid advancement of quantum computing hardware in recent years has been accompanied by the development of a number of quantum computing simulation platforms\footnote{Although not a dedicated quantum computing simulation framework, QuTiP \cite{Johansson2012QuTiP:Systems} is another excellent platform with some quantum information simulation modules.}, including Quipper \cite{Green2013}, IBM Q \cite{IBM-Research2016IBMExperience}, LI$QUi\lvert\rangle$ \cite{Wecker2014Liqui:Computing}, QCL \cite{Omer2005ClassicalProgramming}, Quil \cite{Smith2016AArchitecture}, and Strawberry Fields \cite{Killoran2018StrawberryComputing}. Although the quantum computing simulator ecosystem is now quite large, most, if not all, of these platforms focus on simulating local, noiseless computations performed by a single party. The quantum \emph{network} simulation ecosystem, however, is largely empty.

In this whitepaper, we introduce an open-source framework written in Python \cite{VanRossum2010TheLanguage} and NumPy \cite{Oliphant2007PythonComputing} for simulating quantum networks called the \emph{Simulator for Quantum Networks and Channels}, or \texttt{SQUANCH}. This framework has the following key differences from other similar quantum simulation platforms:

\begin{itemize}
    \item It is designed to simulate quantum networks where parties are connected by noisy quantum channels with configurable error models
    \item It provides well-defined programmatic notions of separate ``agents'', which can manipulate subsets of a distributed quantum state
    \item It is computationally optimized for handling large ``streams'' of quantum information which would be present in sizable quantum networks
    \item Simulations are fully parallelized, with each network node running its own process, mirroring the distributed structure of a quantum network
    \item The framework emphasizes ease of use, imposing minimal syntactical overhead and allowing for rapid development using Python
\end{itemize}

The remainder of this paper is organized as follows. In Section \ref{sec:preliminaries} we give a brief primer on quantum information and introduce quantum agents, channels, and networks. In Section \ref{sec:frameworkOverview}, we present an overview of the \texttt{SQUANCH} framework. We first present the classes that represent quantum information, detailing the \texttt{QSystem}, \texttt{Qubit}, \texttt{QStream}, and \texttt{Gates} modules. Next, we describe the higher-abstraction components which are used to build the nodes and connections of a quantum network: \texttt{Agent}s, \texttt{Channel}s, and \texttt{Errors}. We provide pedagogical demonstrations throughout each of these discussions. Finally, in Section \ref{sec:demos}, we show a series of more sophisticated demonstrations of canonical quantum information protocols implemented with the framework.

\section{Preliminaries}
\label{sec:preliminaries}

In this section, we give a brief primer on quantum computation\footnote{For an excellent in-depth review of this topic, we direct readers to the canonical textbook in quantum computation by Nielsen and Chuang \cite{Nielsen2010QuantumInformation}.}, reviewing some of the concepts necessary to understand the \texttt{SQUANCH} framework. In Section \ref{sec:quantum_networks_channels}, we also provide some basic formalism for describing quantum networks and channels. We encourage readers with familiarity in these topics to skip to Section \ref{sec:frameworkOverview} for an overview of our platform. Throughout this section, a basic knowledge of linear algebra and quantum mechanics is assumed.

\subsection{Qubits and quantum systems}
\label{sec:qubits_and_quantum_systems}
In quantum computation, information is stored in the state of a \emph{qubit}\footnote{There are in fact multiple equivalent models of quantum computation. The prevailing model represents information with qubits, but other models can use discrete $k$-level quantum systems (qudits), or even continuous quantum variables, as detailed in \cite{Killoran2018StrawberryComputing}.}, the quantum generalization of a classical bit. Physically, a qubit can be any two-level quantum system, such as spin or polarization. The state\footnote{What we refer to as the ``state'' is more correctly the projection of the quantum state onto the basis of the observable in question.} of a qubit is a vector $\ket{\psi}$ over $\mathbb{C}^2$ in a linear superposition of states $\ket{\psi}=\alpha \ket{0}+\beta \ket{1}$. The amplitudes $\alpha,\beta\in\mathbb{C}$ obey normalization, such that $\abs{\alpha}^2+\abs{\beta}^2=1$. The state vectors $\{\ket{0},\ket{1}\}$ (physically, the eigenstates of the logical observable) form the \emph{computational basis} which spans the two-dimensional Hilbert space $\mathcal{B}$ containing $\ket{\psi}$. If a \emph{measurement} is performed on the qubit state, the state collapses to $\ket{0}$ or $\ket{1}$, ``returning'' a classical bit $0$ or $1$, with probability $\abs{\alpha}^2$ or $\abs{\beta}^2$, respectively.

A collection of qubits forms a multi-particle \emph{quantum system}. The state of an $N$-qubit quantum system lives in a $2^N$-dimensional Hilbert space $\mathcal{H}_N$ and can be represented by a state vector:
\begin{equation}
    \ket{\Psi}\in\mathcal{H}_N \equiv \bigotimes_{k=N}^1 \mathcal{B}_k,
\end{equation}
where $\bigotimes_{k=N}^1$ denotes a rightward tensor product ordered as $k=N,N-1,\cdots,1$ and $\mathcal{B}_k$ is the state space for each constituent qubit. The system state space $\mathcal{H}_N$ is spanned by the product of the qubit eigenstates $\bigotimes_{k=N}^1 \ket{q_k}$, for $q_k\in\{0,1\}$. These basis vectors are (also) referred to as the computational basis for the $N$-qubit system; they can be written as $\ket{q_1 q_2 \cdots q_N}$ and have the same lexicographic ordering as their binary representations (e.g. $\ket{00},\ket{01},\ket{10},\ket{11}$ for a two-qubit system).

In general, a multi-qubit state $\ket{\Psi}$ cannot be represented as the tensor product of the states of its qubits; that is:
\begin{equation}
    \ket{\Psi} \ne \ket{\psi_1} \otimes \ket{\psi_2} \otimes \cdots \otimes \ket{\psi_N},
\end{equation}
 so the state of each qubit is generally not independent of other qubits. (If $\ket{\psi}$ does take this form, it is called a separable state.) This non-local correlation between quantum states is called entanglement. The massive parallelism of quantum computation is due to superposition and entanglement, as they allow the size of the state space of a quantum system to scale exponentially with the number of qubits. By cleverly exploiting superposition and entanglement, quantum algorithms can achieve polynomial \cite{Grover1996ASearch} or even exponential \cite{Shor1994PolynomialComputer} speedup over classical algorithms for important problems.

\subsection{Open quantum systems}
\label{sec:open_quantum_systems}

The states for quantum systems described in Section \ref{sec:qubits_and_quantum_systems} are \emph{pure} states, which encode all that can be known about a quantum system. However, in many cases, the full quantum state is not accessible to an observer (for example, if the system is a subset of a larger quantum system or if the state is coupled to the environment). In this case, the available quantum state is a statistical mixture of pure states $\ket{\Psi_j}$, each with probability $p_j$, which can be represented by a density matrix $\hat\rho$:
\begin{equation}
    \hat\rho = \sum_j p_j \ketbra{\Psi_j}{\Psi_j}.
\end{equation}

If a quantum system with (pure) state $\ket{\Psi}$ is divided into two subsystems $A$ and $B$, spanned by orthonormal bases $\{\ket{\alpha_i}\}$ and $\{\ket{\beta_j}\}$, respectively, then $\ket{\Psi}=\sum_{ij} \ket{\alpha_i}\ket{\beta_j}$ can equivalently be expressed by the density matrix:
\begin{equation}
    \hat\rho = \ketbra{\Psi}{\Psi}=\sum_{ij} \sum_{i'j'} c_{ij} c_{i'j'}^* \ket{\alpha_i}\ket{\beta_j}\bra{\alpha_{i'}}\bra{\beta_{j'}}.
\end{equation}
If an observer only has access to subsystem $A$, then their (mixed) state $\hat\rho_A$ is given by the partial trace over all inaccessible degrees of freedom (that is, $B$):
\begin{equation}
    \hat\rho_A = \tr_B \hat\rho = \sum_j \bra{\beta_j}\hat\rho\ket{\beta_j}.
\end{equation}

If a measurement of $\hat\rho$ can yield possible outcomes of $\{\ket{j}\}$, then the probability of each outcome is given by $p_j = \tr\left[\hat\rho \ketbra{j}{j}\right]$ and the measurement collapses the state as:
\begin{equation}
\label{eq:density_measurement}
    \hat\rho \mapsto \hat\rho_j = \frac{\ketbra{j}{j} \hat\rho \ketbra{j}{j}^\dagger}{p_{j}}.
\end{equation}

\subsection{Quantum gates}
\label{sec:quantum_gates}

Just as classical algorithms can be represented as a sequence of Boolean logic gates\footnote{More precisely, a Turing machine with arbitrary but finite memory can be constructed from a functionally complete set of logic gates.}, quantum algorithms use quantum gates to manipulate the states of quantum systems. A quantum gate for an $N$-qubit system is a unitary operator $U\in\mathrm{U}(2^N)$; when applied to a quantum state $\hat\rho$, the gate modifies the state as $\hat\rho \mapsto U \hat\rho \,U^\dagger$. 

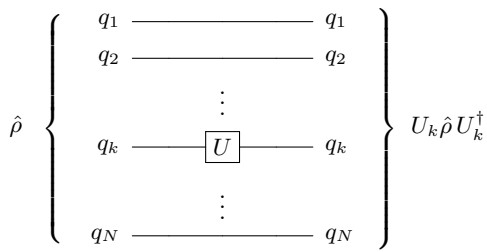
\begin{figure}[h]
\centerline{
\Qcircuit @C=1.5em @R=1.5em {
&   &\lstick{q_1} & \qw & \qw     & \qw   & \rstick{q_1} \qw& \push{\rule{1em}{0em}}\\
&   &\lstick{q_2} & \qw & \qw     & \qw   & \rstick{q_2} \qw& \\
\lstick{\raisebox{-3.2em}{$\hat{\rho}$}\rule{1em}{0em}}& & & & \vdots & & & \rstick{\rule{1.5em}{0em}\raisebox{-3.4em}{$U_k \hat\rho \, U_k^\dagger$}}\\
&   &\lstick{q_k} & \qw & \gate{U}& \qw   & \rstick{q_k} \qw& \\
&   &                   &     & \vdots  &       &                       & \\
&   &\lstick{q_N} & \qw & \qw     & \qw   & \rstick{q_N} \qw& \push{\rule{1em}{0em}}
\gategroup{1}{1}{6}{1}{.5em}{\{}
\gategroup{1}{8}{6}{8}{1em}{\}}
}
}
\caption{Circuit diagram for applying the single-qubit gate $U$ to qubit $q_k$. Wires implicitly denote tensor products with identity.}
\label{fig:single-qubit}
\end{figure}

If a gate acts on a subset of a larger quantum system, their matrix representations are ``padding'' with the identity operator. For example, if $\hat\rho$ represents an $N$-qubit system, and a single-qubit gate $U$ is applied to qubit $k$, then the state transforms as:
\begin{equation}
\label{eq:identity_padding}
    \hat\rho \mapsto U_k \hat\rho \, U_k^\dagger \,:\, U_k \equiv \id_N \otimes \cdots \otimes U \otimes \id_{k-1} \otimes \cdots \otimes \id_1,
\end{equation}
where $\id_j$ denotes identity applied to the $j$th qubit. Quantum gates are commonly described pictographically with quantum circuit diagrams, such as the one in Figure \ref{fig:single-qubit}, which depicts the transformation in Equation \ref{eq:identity_padding}. A list of circuit symbols for built-in gates included in \texttt{SQUANCH} is provided in Appendix \ref{sec:gatesList}.

It is straightforward to define many (and in fact, using \texttt{SWAP}, all) multi-qubit gates using a similar strategy as in Equation \ref{eq:identity_padding}. For example, the controlled-X or \texttt{CNOT} gate, common for constructing entanglement in a quantum system, has a matrix representation for a two-qubit system:
\begin{equation}
    \texttt{CNOT} \equiv 
    \begin{scriptsize}
    \begin{pmatrix}
    1 & 0 & 0 & 0 \\
    0 & 1 & 0 & 0 \\ 
    0 & 0 & 0 & 1 \\
    0 & 0 & 1 & 0 \\
    \end{pmatrix}
    \end{scriptsize}.
\end{equation}
Equivalently, we can define the gate $\texttt{CNOT}_{jk}$ acting on control qubit $j$ and target qubit $k$ in an $N$-qubit system as:
\begin{equation}
     \texttt{CNOT}_{jk} \equiv \ketbra{0}{0}_j \otimes\id_k + \ketbra{1}{1}_j \otimes\sigma_{x;k},
\end{equation}
where $\sigma_{x;j}$ is Pauli-X applied to qubit $j$ and padding with identity is implicit.\footnote{For the remainder of this paper, we adopt the convention that a $k$-qubit operator $\Omega$ acting on qubits $i_1\cdots i_k$ is notated $\Omega_{i_1\cdots i_k}$ and implicitly includes ordered tensor products with $\id_j$ for $j\not\in\{i_1\cdots i_k\}$.}

\subsection{Quantum agents}
\label{sec:quantum_agents}
Algorithms for transmitting and manipulating quantum information are frequently explained in terms of a colorful cast of characters starring Alice and Bob. In the \texttt{SQUANCH} framework, these ``quantum actors'' are represented by \texttt{Agent}s. An agent has a set of classical and quantum channels connecting it to other agents, classical memory, which can store arbitrary data, and quantum memory, which is an array of qubits. It also has runtime logic in the form of a classical program with access to operations that can manipulate quantum information, such as gates and measurement. For completeness, a formal definition of a quantum agent is given in Appendix \ref{sec:quantum_agent_definition}. 

\subsection{Quantum networks and channels}
\label{sec:quantum_networks_channels}

In this section, we provide some formalism to define quantum channels and quantum networks as used in our simulation framework.

Conceptually, a \emph{quantum channel} is a communication channel which transmits quantum information (qubits) and an associated noise model\footnote{Most common definitions of a quantum channel define the channel as the noise model $\mathcal N$ itself; as before, we make the distinction to mirror the codebase structure.} $\mathcal N$ which modifies the qubits, generally through coupling to the environment. Mathematically, $\mathcal N$ is a linear completely-positive trace-preserving map which acts on a composite state consisting of the input quantum state state $\hat\rho_\text{in}$ and the environment $\hat\xi_\text{in}$ (assumed by convention to be in the initial state $\ketbra{0}{0}$). \cite{Gyongyosi2012PropertiesChannel}

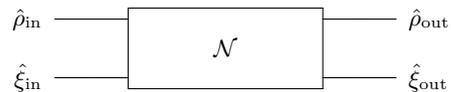
\begin{figure}[ht]
\centerline{
\Qcircuit @C=1.5em @R=1.5em {
\lstick{\mbox{$\hat\rho_\text{in}$}} & \qw & \multigate{1}{\rule{3em}{0em}\mathcal{N}\rule{3em}{0em}} & \qw & \rstick{\mbox{$\hat\rho_\text{out}$}} \qw \\
\lstick{\mbox{$\hat\xi_\text{in}$}} & \qw & \ghost{\rule{3em}{0em}\mathcal{N}\rule{3em}{0em}} & \qw & \rstick{\mbox{$\hat\xi_\text{out}$}} \qw
}
}
\caption{Circuit diagram representing a quantum channel $\mathcal{N}$. Frequently, the coupling of the channel to the environment is assumed and not explicitly drawn, as in Figure \ref{fig:shor_code}.}
\end{figure}

While qubits are in the channel, the state of the composite system undergoes unitary evolution as $\hat\rho_\text{in} \otimes \hat\xi_\text{in} \mapsto U \left(\hat\rho_\text{in} \otimes \hat\xi_\text{in}\right) U^\dagger$, which can allow information to ``leak'' from the system $\hat\rho$ to the environment $\hat\xi$. After the qubit is retrieved from the channel, the output state $\hat\rho_\text{out}$ is given by tracing over inaccessible environmental degrees of freedom \cite{Gyongyosi2012PropertiesChannel}:

\begin{equation}
    \hat\rho_\text{out} = \mathcal{N}\left(\hat\rho_\text{in}\right) = \tr_\xi \left[U \left(\hat\rho_\text{in} \otimes \hat\xi_\text{in}\right) U^\dagger\right].
\end{equation}

We direct readers to papers by Gyongyosi, et al., for in-depth discussions of the properties \cite{Gyongyosi2012PropertiesChannel} and capacities \cite{Gyongyosi2018ACapacities} of quantum channels.


Finally, a \emph{quantum network} can be defined\footnote{Quantum networks generally lack the standard definition that quantum computers or quantum channels have. We provide a formal definition here for completeness, but it should not be interpreted as authoritative.} as a directed graph $N=(A,C)$, where $A$ is a set of agents and $C$ is a set of quantum and classical channels\footnote{Since our definition of a quantum agent includes a list of channels, explicitly including $C$ in the definition is redundant but adds clarity.}. Other practical network components, such as quantum repeaters \cite{Azuma2015All-photonicRepeaters}, can be modeled (both mathematically and programmatically) as quantum agents.

\section{Framework overview}
\label{sec:frameworkOverview}

\begin{figure*}[t]
    \includegraphics[width=\textwidth]{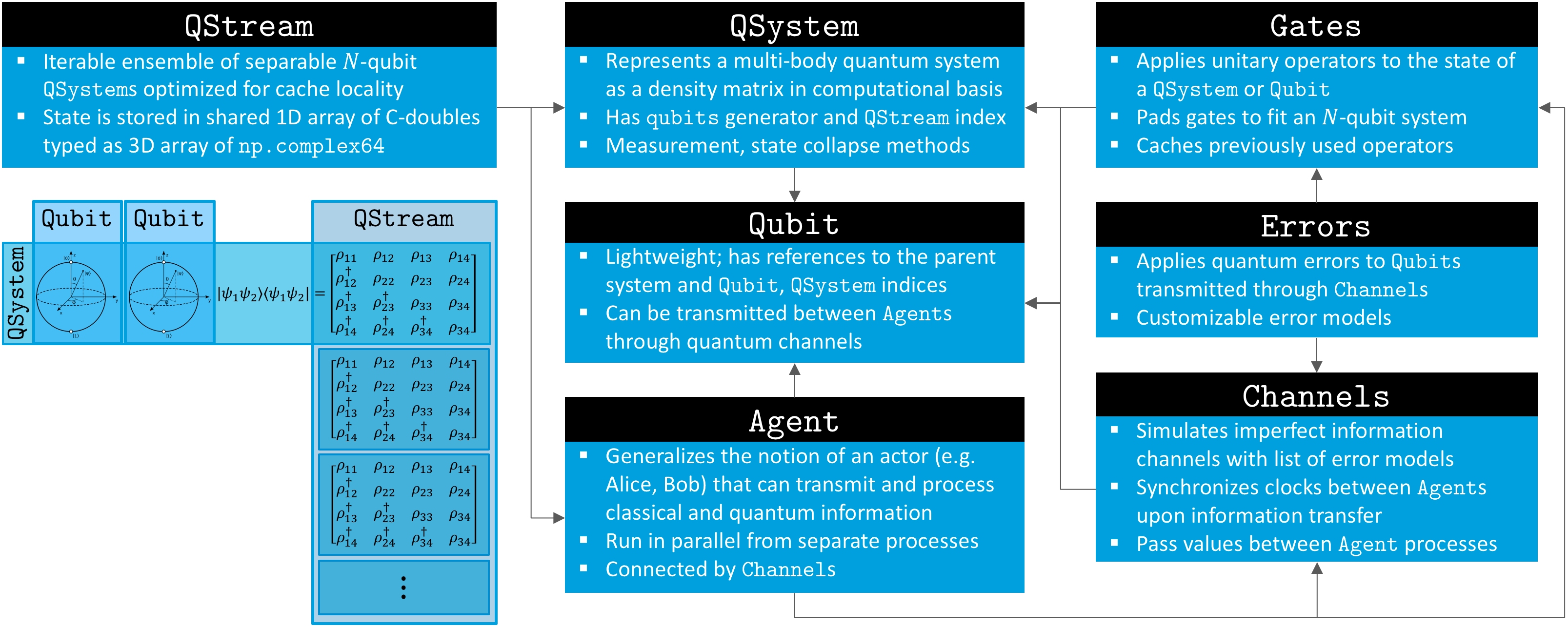}
    \caption{Schematic overview of the modules available in the \texttt{SQUANCH} framework, with an illustration in the lower left showing how quantum streams and states are represented. The \texttt{QSystem} is the most fundamental class, representing a multi-particle quantum state as a density matrix. Ensembles of quantum systems are efficiently handled by \texttt{QStream}s, and each \texttt{QSystem} has references to its constituent \texttt{Qubit}s. Functions in the \texttt{Gates} module can be used to manipulate the state of a quantum system. \texttt{Agent}s are generalized quantum-mechanical ``actors'' which are initialized from a \texttt{QStream} instance and can alter the state of the quantum systems in their stream object, typically by interacting directly with \texttt{Qubit}s. \texttt{Agent}s run in parallel from separate processes and are connected by quantum and classical \texttt{Channel}s, which apply customizable \texttt{Error} models to the transmitted information and synchronize agent clocks.}
    \label{fig:overview}
\end{figure*} 

\texttt{SQUANCH}\footnote{This manuscript refers to \texttt{SQUANCH} version 1.1.0, available at \url{github.com/att-innovate/squanch} or from the Python Package Index using \texttt{pip install squanch}. Documentation is available at \url{att-innovate.github.io/squanch}.} is a Python-based framework for simulating the dynamics of quantum networks and provides an intuitive, object-oriented API for representing and manipulating ensembles of separable quantum systems. A schematic overview of the \texttt{SQUANCH} framework is depicted in Figure \ref{fig:overview}. In this section, we describe the structure of the framework, detailing the modules to represent and manipulate quantum information and to define and simulate quantum networks. Full documentation is available online, but an abridged API reference is available at the end of this manuscript in Appendix \ref{sec:api}.

\subsection{Working with quantum information}

\subsubsection{\texttt{QSystem}s and \texttt{Qubit}s}

The most fundamental class in the framework is the \texttt{QSystem}, which represents the state of a multi-body, maximally entangleable quantum system. The state of each qubit $\ket{\psi_k}$ of a quantum system is represented in the abstract computational basis\footnote{See Section \ref{sec:qubits_and_quantum_systems} for more detail.} $\mathcal{B}_k$ spanned by $\left\{\ket{0}_k, \ket{1}_k\right\}$, and the state of the system as a whole is represented by a density matrix, which is implemented in memory as a NumPy array of \texttt{np.complex64} values. 

A \texttt{QSystem} contains a \texttt{QSystem.qubits} generator,\footnote{Generators are used instead of lists for performance reasons, although they do introduce a small number of peculiarities in the code, which are discussed in the documentation.} which enumerates the qubits of the quantum system. Each qubit is represented by a \texttt{Qubit} object, a lightweight wrapper class containing only a reference to the parent \texttt{QSystem} and the index of the qubit within the system\footnote{If the parent system is part of a \texttt{QStream}, the \texttt{Qubit} will also contain an index for the location of the system within the stream.}. 

\texttt{Qubit} and \texttt{QSystem} objects contain methods for simulating qubit measurement in the computational basis, as described in Sections \ref{sec:qubits_and_quantum_systems} and \ref{sec:open_quantum_systems}. \texttt{Qubit.measure()} calls \texttt{QSystem.measure\_qubit(index)}, which returns a bit value for the result of the measurement. Measurement of qubit $k$ partially collapses the system state $\hat\rho$ by choosing a random outcome $\ket{\psi_k}=\ket{i}$ with probability $p_{i;k} = \tr \left[\hat\rho \left(\ketbra{i}{i}\right)_k\right]$ for $i\in\{0,1\}$, modifying the state according to Equation \ref{eq:density_measurement}.

\texttt{QSystem}s also contain a method, \texttt{QSystem.apply()}, for applying an $N$-dimensional unitary operator to modify the system state. In practice, this method is rarely used, as qubits can be more intuitively manipulated using functions in the \texttt{gates} module, which provides a set of common single- and multi-qubit quantum logic gates.

\subsubsection{Quantum gates}

Quantum gates are implemented as standard Python functions which take one or more \texttt{Qubit}s as arguments, return nothing, and call \texttt{Qubit.apply()} or \texttt{QSystem.apply()} to apply a unitary operator to the quantum state. A list of natively-included gates is provided in Appendix \ref{sec:gatesList}. The corresponding $N$-qubit operator (as described in Section \ref{sec:quantum_gates}) for each gate is lazily evaluated, cached, and applied directly to the the parent \texttt{QSystem.state}, modifying its density matrix in-place. A simple example of state preparation and measurement using built-in gate functions is demonstrated\footnote{For the remainder of this paper, we assume that all programs are run with the standard import statements of \texttt{import numpy as np}, \texttt{import matplotlib.pyplot as plt}, and \texttt{from squanch import *}.} in Program \ref{prg:qsys_demo}.

\begin{code}
    \pythonfile{code/qsys_demo.py}
    \caption{Preparing and measuring the entangled state $\ket{+}=\frac{1}{\sqrt{2}}\left(\ket{00}+\ket{11}\right)$. The measurement results \texttt{bit1} and \texttt{bit2} will always be equal.}
    \label{prg:qsys_demo}
\end{code}

\begin{figure}[h]
\mbox{
\scriptsize
\Qcircuit @C=.3em @R=.7em {
  \lstick{q_0}	& 	\gate{H}    &   \gate{\phi_2}   &   \gate{\phi_3}   &   \qw      &   \gate{\phi_N}   &   \qw        &   \qw           &   \qw      &   \qw         &   \qw         &   \qw     &   \qw           &   \qw        &   \rstick{y_{N-1}} \qw       \\
  \lstick{q_1}   &   \qw         &   \ctrl{-1}     &   \qw           &   \qw      &   \qw           &   \gate{H}   &   \gate{\phi_2}   &   \qw       & \gate{\phi_{N-1}}  &   \qw         &   \qw     &   \qw           &   \qw        &   \rstick{y_{N-2}} \qw       \\
  \lstick{q_2}   &   \qw         &   \qw           &   \ctrl{-2}     &   \qw      &  \qw            &   \qw        &   \ctrl{-1}     &   \qw      &   \qw         &   \gate{H}   &   \qw     &   \gate{\phi_{N-2}}   &   \qw        &   \rstick{y_{N-3}} \qw       \\
  \lstick{\vdots }     &             &                 &                 &   \ddots   &                 &              &                 &   \ddots   &           &              &   \ddots  &                 &              &   \rstick{\vdots }             \\
  \lstick{} &   &   &     &    &     &  &   &     &     &   &     &  &   & \rstick{} \\
  \lstick{q_{N-1}} &   \qw       &   \qw           &   \qw           &   \qw      &   \ctrl{-5}     &    \qw       &   \qw           &   \qw      &   \ctrl{-4}     &   \qw        &   \qw     &   \ctrl{-3}     &   \gate{H}   &   \rstick{y_{0}} \qw
}
}
\caption{Circuit implementation of a quantum Fourier transform. $\phi_m$ denotes the phase gate $\smat{1}{0}{0}{\omega_m}$, where $\omega_m\equiv e^{\frac{2 \pi i}{2^m}}$ is a primitive root of unity.}
\label{fig:qft_circuit}
\end{figure}
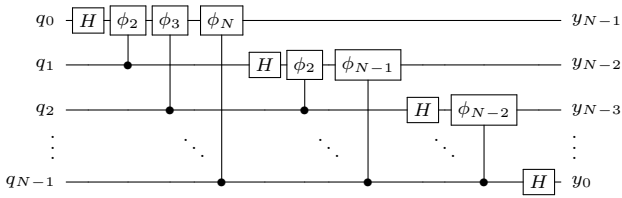

Additional gates can be added through compositions of existing gates, which can approximate any unitary operator with arbitrary accuracy \cite{Barenco1995ElementaryComputation}, or by directly writing a function to implement the gate operation. As an example, consider the quantum Fourier transform \cite{Nielsen2010QuantumInformation}, which operates on the $2^N$ lexicographically-enumerated $N$-qubit basis states $\{\ket{j}, \ket{k} \in \bigotimes_{k=N}^1 \mathcal{B}_k \}$ as:
\begin{equation}
    \sum_j \alpha_j \ket{j} \mapsto \sum_k \left( \frac{1}{\sqrt{2^N}}\sum_{n=0}^{2^N-1} e^\frac{2\pi i k n}{2^N} \alpha_n \right) \ket{k}.
\end{equation}
This operation can be implemented iteratively with the quantum circuit depicted in Figure \ref{fig:qft_circuit}.

Programmatically implementing this relatively complicated quantum circuit is straightforward in \texttt{SQUANCH}, as demonstrated in Program \ref{prg:qft_demo}.

\begin{code}
    \pythonfile{code/qft_demo.py}
    \caption{Implementing the quantum Fourier transform using built-in gates. The \texttt{CPHASE} operator is the controlled-phase gate, which acts as the identity if the control qubit is in state $\ket{0}$ and maps $\ket{0}\rightarrow\ket{0}$ and $\ket{1}\rightarrow e^{i\phi}\ket{1}$ if the control qubit is in state $\ket{1}$.}
    \label{prg:qft_demo}
\end{code}

\subsubsection{\texttt{QStream}s -- performant quantum ensembles}

An ensemble of separable quantum systems, such as one million Bell pairs, is represented by a ``quantum stream'', or \texttt{QStream}, which stores the collective state of its \texttt{QSystem}s in a contiguous block of pre-allocated memory to optimize cache locality. A \texttt{QStream} is instantiated from two dimensions: the number of qubits in each quantum system, and the number of systems in the stream. The \texttt{QStream} is an iterable Python object, and most simulations will contain an iteration over the quantum systems in the stream. To minimize overhead, \texttt{QSystems} can be quickly instantiated from an existing section of the \texttt{QStream.state} array.

By default \texttt{QStream} state is stored in a shared memory as a C-type array of \texttt{double}s, which is type-casted as a 3D array of \texttt{np.complex64} values. Storing the state in shared memory allows multiple agents to work in parallel. For performance reasons, there is no explicit concurrency safety when a system is modified by multiple agents, as sending and receiving \texttt{Qubits} are blocking operations which allow for naturally safe parallelism.

\subsection{Simulating quantum networks}


The top-level classes that provide the greatest abstraction are \texttt{Agent}s and \texttt{Channel}s, which implement the nodes and connections in a quantum network, respectively. 

\subsubsection{\texttt{Agent}s -- Alice and Bob in code}

An \texttt{Agent} generalizes the notion of an actor (e.g. Alice, Bob) that can send, receive, store, and manipulate classical and quantum information. We provide a formal definition for a quantum agent in Section \ref{sec:quantum_agents}; the structure of the \texttt{Agent} class mirrors this definition. 

\texttt{Agent}s are instantiated from a \texttt{QStream} object, and each \texttt{Agent} has a classical memory, which can store arbitrary Python objects, a quantum memory, which stores incoming qubits and can be modified to simulate error models. \texttt{Agents} have a processor in the form of a \texttt{run()} method that implements runtime logic. Additionally, \texttt{Agent}s maintain internal clocks which are updated when sending or receiving information to each other; this allows users to roughly quantify the performance of various networking protocols in terms of simulated elapsed time. During simulations, \texttt{Agent}s run in parallel from separate processes, synchronizing clocks and passing information between each other through \texttt{Channels} using \texttt{qsend/qrecv} and \texttt{csend/crecv}.

A typical workflow for creating a quantum network simulation using \texttt{Agent}s is:

\begin{enumerate}

    \item Define the runtime logic for each agent class. This usually involves an iteration over \texttt{self.qstream}, calls to send and receive information to/from other agents, and a call to output the results of the simulation to the main process using \texttt{self.output(<objects>)}.
    
    \begin{itemize}
    
        \item If the agent is a ``sender'', the runtime logic typically iterates \texttt{for qsystem in self.qstream}, followed by procedures to manipulate and transmit the qubits of \texttt{qsystem}.
        
        \item If the agent is a ``receiver'', the runtime logic is usually \texttt{for \_ in self.qstream: qubit = self.qrecv(<other\_agent>)}, followed by processing and output procedures.
        
        \item Of course, mixtures of both formats may be used.
        
    \end{itemize} 
    
    \item Create and format a \texttt{QStream} object. State preparation may be done as part of the agent logic or in the main process.
    
    \item Create an output dictionary (e.g. \texttt{out = Agent.shared\_output()}) and make instances for each agent class from the quantum stream (e.g. \texttt{alice = Alice(qstream, out)}).
    
    \item Define the structure of the network by connecting agent instances using \texttt{alice.qconnect(bob)} and/or \texttt{alice.cconnect(bob)}, optionally specifying the channel model(s) to use.
    
    \item Run the simulation using \texttt{Simulation(alice, bob, ...).run()} and post-process and visualize the output data. Output results for each agent will be returned to corresponding value in the previously created \texttt{out} dictionary (e.g. \texttt{alice\_out = out["Alice"]}).
    
\end{enumerate}
This general program structure is demonstrated in Program \ref{prg:agent_demo}.

\begin{code}
	\pythonfile{code/agent_demo.py}
	\caption{A simple quantum network ``ping'' protocol implemented using agents. Alice prepares single-qubit states to send to Bob, who measures them and replies with the outcomes.}
	\label{prg:agent_demo}
\end{code}

\subsubsection{\texttt{Channel}s -- imperfect communication lines}

Classical and quantum channels, which are derived from the \texttt{CChannel} and \texttt{QChannel} base classes, represent noisy information channels physically connecting distant nodes in a network. A formal definition of a quantum channel is provide in Section \ref{sec:quantum_networks_channels}; a practical example would be a fiber optic line with a specified length and attenuation coefficient. Functionally, \texttt{Channel}s are multiprocessed queues that allow routines to communicate between processes. \texttt{Channel}s track transmission times and speed-of-light delays (through an optional \texttt{length} parameter which can specify the physical size of the channel in kilometers) and apply customizable error models to the transmitted information. Quantum error models are implemented by extending the base \texttt{QError} class, which maintains a reference to the parent channel and has a single \texttt{apply(qubit)} method to apply the error model to a transmitted qubit. An implementation of a simple quantum error model is shown in Program \ref{prg:channel_demo}.

\begin{code}
	\pythonfile{code/channel_demo.py}
	\caption{An extension of the quantum ``ping'' protocol from Program \ref{prg:agent_demo} which adds simple quantum error models to the network. (\texttt{Alice} and \texttt{Bob} are defined as before.) Agent timing functionality is also demonstrated given a specified channel length of 1km.}
	\label{prg:channel_demo}
\end{code}

\section{Demonstrations}
\label{sec:demos}

Finally, in this section we present several demonstrations of canonical experiments and protocols involving quantum information. For brevity, we omit code for plotting or displaying images and assume standard imports. The \texttt{SQUANCH} documentation website (see Appendix \ref{sec:sourceCode}) contains the full source code for each demonstration, along with more detailed step-by-step explanations.

\subsection{Quantum teleportation}

Quantum teleportation allows two parties that share an entangled pair to transfer an arbitrary quantum state using only classical communication \cite{Bennett1993TeleportingChannels}. This process has tremendous applicability to quantum networks, transferring fragile quantum states between distant nodes. Conceptually, quantum teleportation is the inverse of superdense coding.

In general, all quantum teleportation experiments have the same underlying structure. Two distant parties, Alice and Bob, are connected via a classical information channel and share a maximally entangled state. Alice has an unknown state $\ket{\psi}$ which she wishes to send to Bob. She performs a joint projective measurement of her state and her half of the entangled state and communicates the outcomes to Bob, who operates on his half of the entangled state accordingly to reconstruct $\ket{\psi}$. In this demo, we’ll implement the canonical two-party quantum teleportation protocol:

\begin{enumerate}
    \item Alice generates an entangled two-particle state $\ket{AB} = \frac{1}{\sqrt{2}}\left(\ket{00}+\ket{11}\right)$, keeping half of the state and sending the other half to Bob.
    \item Alice entangles her qubit $\ket{\psi}$ with her ancilla $\ket{A}$ by applying controlled-not and Hadamard operators.
    \item Alice jointly measures $\ket{\psi}$ and $\ket{A}$ and communicates the outcomes to Bob through a classical channel. Bob’s qubit is now in one of four possible Bell states, one of which is $\ket{\psi}$, and he will use Alice’s two bits to recover $\ket{\psi}$.
    \item Bob applies a Pauli-X operator to his qubit if Alice’s ancilla $A$ collapsed to $\ket{1}$, and he applies a Pauli-Z operator to his qubit if her state $\ket{\psi}$ collapsed to $\ket{1}$. He has thus transformed $\ket{B}\mapsto\ket{\psi}$.
\end{enumerate}

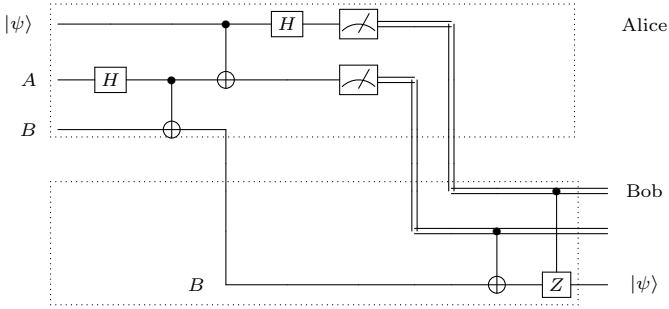
\begin{figure}[h]
\mbox{
\scriptsize
\Qcircuit @C=1.75em @R=1.2em {
\lstick{\ket{\psi} \rule{0.5em}{0em}} & \qw & \qw &\ctrl{1}&\gate{H}&\meter&\cw  &\cw \cwx[1]&& \push{\rule{0em}{1.2em}}& &\mbox{Alice}\\
\lstick{A\rule{0.5em}{0em}} & \gate{H} & \ctrl{1} &\targ& \qw &\meter  &\cw \cwx[1]&\cwx[1] &&& \push{\rule{0em}{1.5em}} &\\
\lstick{B\rule{0.5em}{0em}} & \qw & \targ & \qw \qwx[1] & & & \cwx[1] & \cwx[1] && \push{\rule{1em}{0em}} && \\ 
&& & \qwx[1] &  &&\cwx[1] &\cwx[1] &&&  & \\
&& & \qwx[1] &  &&\cwx[1] &  &\cw&\control \cw  \qwx[2]& \cw & \mbox{Bob}\\
& \push{\rule{0em}{1.2em}} & & \qwx[1] &  &&  &\cw  &\control \cw  \qwx[1]&\cw& \cw &\\
&& &\lstick{B\rule{0.5em}{0em}}   &\qw  &\qw&\qw  &\qw  &\targ &\gate{Z} & \push{\rule{0em}{1.5em}} \qw&\ket{\psi}
\gategroup{1}{1}{3}{10}{.7em}{.}
\gategroup{5}{1}{7}{10}{.7em}{.}
}
}
\caption{Circuit diagram for the two-party quantum teleportation simulation described above.}
\label{fig:quantum_teleportation_circuit}
\end{figure}

This protocol is illustrated in the circuit diagram shown in Figure \ref{fig:quantum_teleportation_circuit}.

\begin{figure}[t]
    \centering
    \includegraphics[width=\columnwidth]{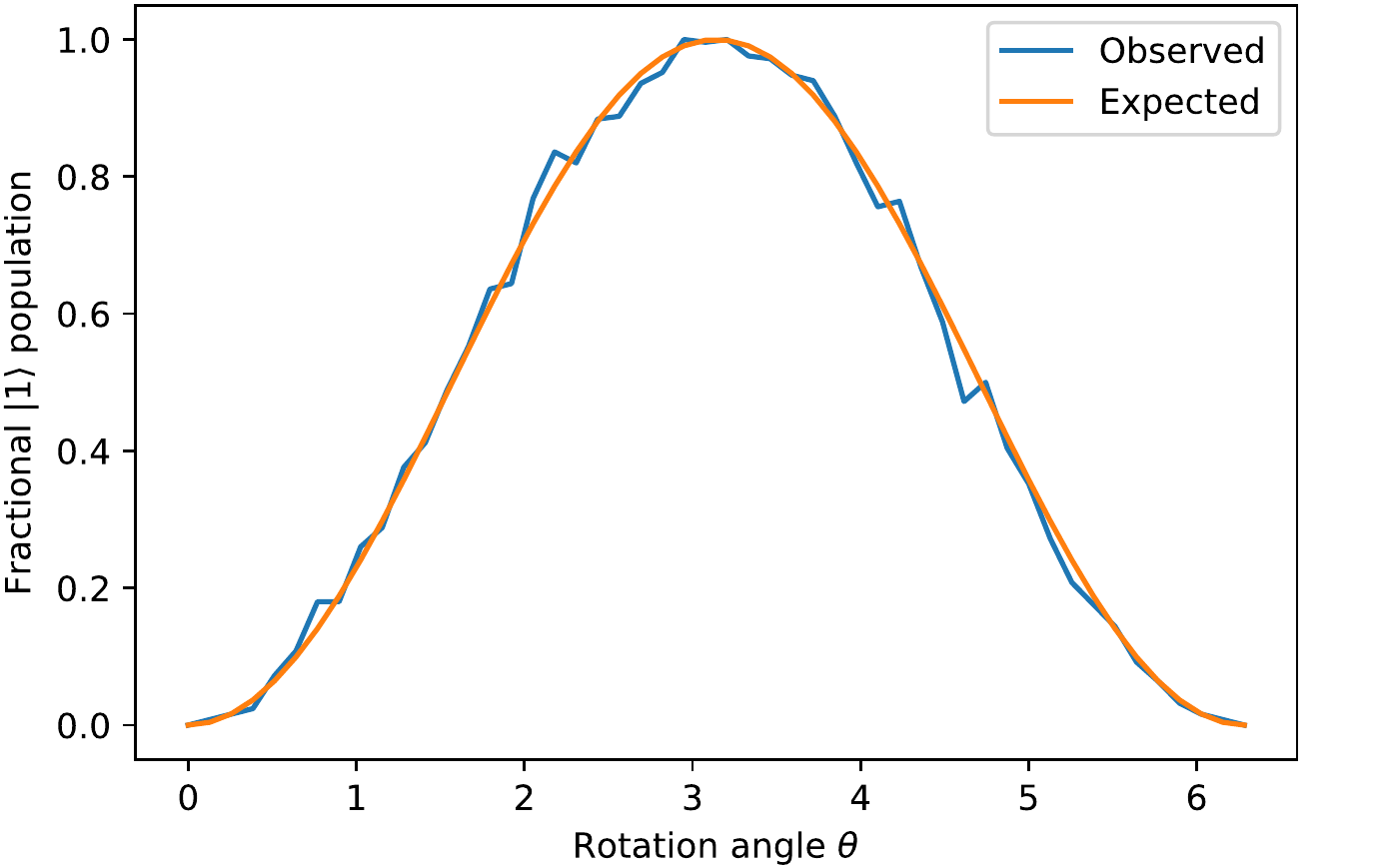}
    \caption{Observed and expected fractional populations of measurements resulting in $\ket{\psi}\rightarrow\ket{1}$ for the two-party quantum teleportation simulation with an ensemble size of 250.}
    \label{fig:teleportation_results}
\end{figure}

For this demonstration, we'll prepare an ensemble of qubits in the state $\ket{q_\theta} = R_X (\theta)\ket{0}$ for various values of $\theta\in[0,2\pi]$ and compare the expected and observed outcomes. (See Appendix \ref{sec:gatesList} for the definition of $R_X (\theta)$.) The results of the simulated experiment are shown in Figure \ref{fig:teleportation_results}.

\begin{code}
	\pythonfile{code/quantum_teleportation.py}
	\caption{Implementation of a two-party quantum teleportation experiment using \texttt{SQUANCH}.}
\end{code}

\subsection{Superdense coding}

Superdense coding is a process whereby two parties connected via a quantum channel and sharing an entangled pair can send two classical bits of information using only a single qubit \cite{Bennett1992CommunicationStates}. Conceptually, superdense coding is the inverse of quantum teleportation. 

\begin{figure}[h]
\scriptsize
\mbox{
\Qcircuit @C=1.75em @R=1.2em {
& \push{\rule{1.5em}{0em}}& & &			& 	 						&  &  &  &&&\mbox{Alice}\\
\lstick{b_1}& \cw & \cw & \cw &	\cw  					&\control \cw \qwx[2]&\cw &\cw  &\cw  &\cw&\cw&  \\
\lstick{b_2}& \cw & \cw & \cw &	\control \cw \qwx[1]	&\cw  &\cw   &\cw  &\cw  &\cw&\rstick{\rule{0.5em}{0em}} \cw&  \\
& & & \qwx[1]&\targ&\gate{Z}&\qw \qwx[1]&  &  &&\push{\rule{0em}{1.5em}}&\\
\lstick{\ket{0}}& \gate{H} & \ctrl{1}& \qw  &\mbox{Charlie}  &  &\qwx[1] & & & &&\\
\lstick{\ket{0}}& \push{\rule{0em}{1em}} \qw & \targ & \qw\qwx[1] &  &  &\qwx[1] &  &   &&&\\
& & & &  &  &\qwx[1] &  &  &&&\mbox{Bob}  \\
& & & \push{\rule{0em}{1.5em}}&  &  &  &\ctrl{1}&\gate{H}&\meter&\cw&\mbox{$b_1$}  \\
& & & \qwx[-3] &\qw  &\qw  &\qw  &\targ&\qw  &\meter&\push{\rule{0em}{1.5em}}\cw &\mbox{$b_2$}
\gategroup{1}{2}{4}{11}{.7em}{.}
\gategroup{7}{4}{9}{11}{.7em}{.}
\gategroup{5}{2}{6}{4}{.7em}{.}
}
}
\caption{Circuit diagram for the three-party quantum superdense coding experiment described above.}
\label{fig:superdense_coding_circuit}
\end{figure}
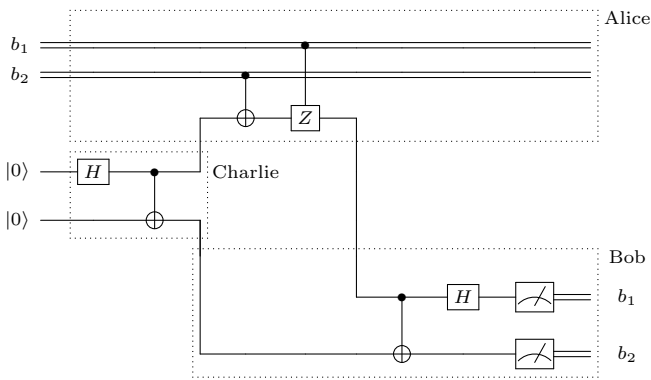

In this demonstration, we'll implement the three-party superdense coding protocol depicted in the circuit diagram in Figure \ref{fig:superdense_coding_circuit}. Charlie distributes entangled particles to Alice and Bob. Alice encodes her information in her particles and sends them to Bob, who decodes the information by matching Alice’s qubits with his half of the shared state received from Charlie. More precisely:

\begin{enumerate}
    \item Charlie generates entangled pairs in the state $\ket{AB} = \frac{1}{\sqrt{2}}\left(\ket{00}+\ket{11}\right)$ and sends one particle to Alice and the other to Bob.
    \item Alice has a number of bits she wishes to send to Bob. For each pair, she encodes her two bits $b_1$ and $b_2$ in the relative sign and phase of her qubit by conditionally applying $\sigma_z$ and $\sigma_x$, respectively. She then sends the modified qubit to Bob.
    \item Bob disentangles the X and Z components of the qubit by applying $\mathrm{CNOT}$ and $H$ to the qubits he received from Alice and from Charlie. He then measures each qubit to obtain $b_1$ and $b_2$, respectively.
\end{enumerate}

For this demonstration, Alice will send data to Bob in the form of a serialized bitstream representing an image. We'll use the built-in timing functionality to track the simulated time for each agent to complete their part of the protocol. Since superdense coding could be used as a networking protocol in the foreseeable future, even very rudimentary simulated timing data could be useful to quantify the performance of the algorithm, especially if data validation and error correction through multiple transmission attempts is simulated. We assume a photon pulse interval of 1ns and a spatial separation between Alice and Bob of 1km, with Charlie at the midpoint. All agents are connected with the \texttt{FiberOpticQChannel} model, which simulates $0.16\frac{\mathrm{dB}}{\mathrm{km}}$ attenuation errors by randomly changing transmitted \texttt{Qubit}s to \texttt{None}.

\begin{code}
	\pythonfile{code/superdense_coding.py}
	\caption{Simulation of transmitting a serialized bitstream of an image via superdense coding.}
	\label{prg:superdense_coding_demo}
\end{code}

\begin{figure}[h]
    \centering
    \includegraphics[width=\columnwidth]{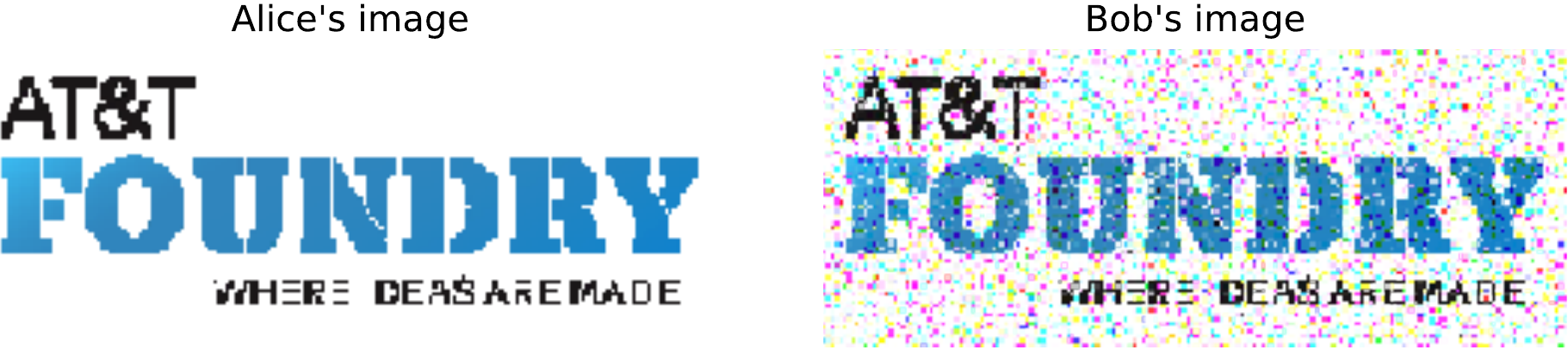}
    \caption{Results of Program \ref{prg:superdense_coding_demo}, showing Alice's original image (left), and Bob's received image (right). Bob applies no correction for attenuation errors, replacing dropped qubits with bit pairs of $(0,0)$.}
    \label{fig:superdense_coding_results}
\end{figure}

\begin{figure*}[t]
    \centering
    \includegraphics[width=\textwidth]{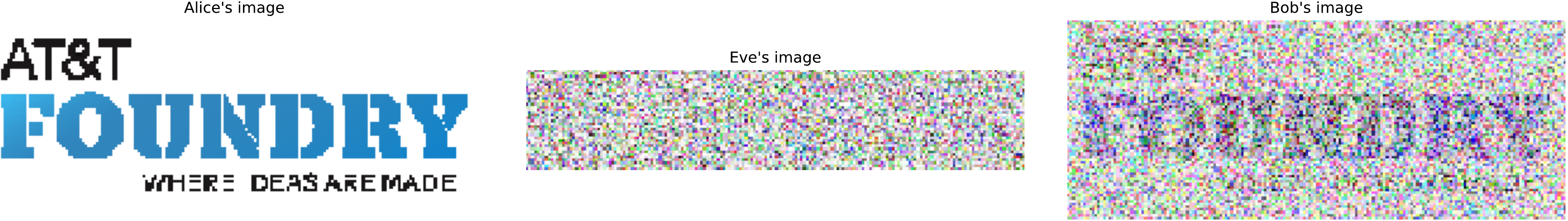}
    \caption{The results of the simulated interception attack simulated by Program \ref{prg:man_in_the_middle_demo}. Since Eve only has access to half of the entangled $\ket{AB}$ state, she recovers only random noise, and Bob's image is half-corrupted, alerting him to the presence of an eavesdropper.}
    \label{fig:man_in_the_middle_results}
\end{figure*}

\subsection{Man-in-the-middle attack}

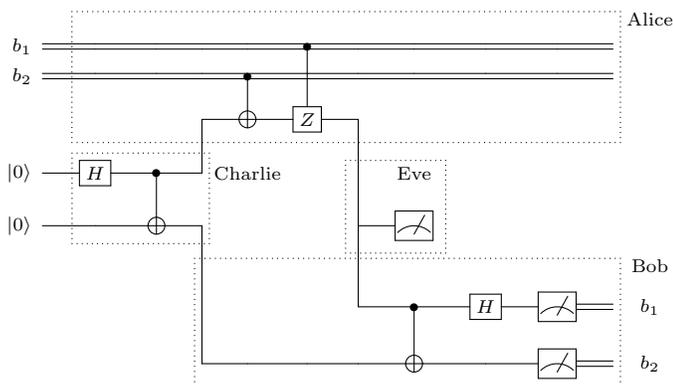
\begin{figure}[H]
\scriptsize
\mbox{
\Qcircuit @C=1.75em @R=1.2em {
& \push{\rule{1.5em}{0em}}& & &			& 	 						&  &  &  &&&\mbox{Alice}\\
\lstick{b_1}& \cw & \cw & \cw &	\cw  					&\control \cw \qwx[2]&\cw &\cw  &\cw  &\cw&\cw&  \\
\lstick{b_2}& \cw & \cw & \cw &	\control \cw \qwx[1]	&\cw  &\cw   &\cw  &\cw  &\cw&\rstick{\rule{0.5em}{0em}} \cw&  \\
& & & \qwx[1]&\targ&\gate{Z}&\qw \qwx[1]&  &  &&\push{\rule{0em}{1.5em}}&\\
\lstick{\ket{0}}& \gate{H} & \ctrl{1}& \qw  &\mbox{Charlie}  &  &\qwx[1] &  \mbox{Eve}& & &&\\
\lstick{\ket{0}}& \push{\rule{0em}{1em}} \qw & \targ & \qw\qwx[1] &  &  &\qwx[1] &  \meter &   &&&\\
& & & &  &  &\qwx[1] &  &  &&&\mbox{Bob}  \\
& & & \push{\rule{0em}{1.5em}}&  &  &  &\ctrl{1}&\gate{H}&\meter&\cw&\mbox{$b_1$}  \\
& & & \qwx[-3] &\qw  &\qw  &\qw  &\targ&\qw  &\meter&\push{\rule{0em}{1.5em}}\cw &\mbox{$b_2$}
\gategroup{1}{2}{4}{11}{.7em}{.}
\gategroup{7}{4}{9}{11}{.7em}{.}
\gategroup{5}{2}{6}{4}{.7em}{.}
\gategroup{5}{7}{6}{8}{1.2em}{.}
}
}
\caption{Circuit diagram demonstrating a naive interception attack on the superdense networking protocol demonstrated in Figure \ref{fig:superdense_coding_circuit}. Eve's meter represents measurement of a qubit and transmission of a new qubit in the observed state.}
\label{fig:man_in_the_middle_circuit}
\end{figure}

In this demo, we show how quantum networks can be resistant to interception (``man-in-the-middle'') attacks by using a modified version of the superdense coding demonstration presented in Program \ref{prg:superdense_coding_demo}. As in the previous demo, Charlie will distribute Bell pairs to Alice and Bob, and Alice will attempt to send a classical message to Bob. However, a fourth party, Eve, will try to naively intercept the message Alice sends to Bob. Eve will measure each qubit from Alice, record the result, and re-transmit the qubit to Bob. This scenario is illustrated in the circuit diagram shown in Figure \ref{fig:man_in_the_middle_circuit}.

This four-party quantum network scenario is simulated below in Program \ref{prg:man_in_the_middle_demo}.

\begin{code}
	\pythonfile{code/man_in_the_middle.py}
	\caption{A four-party interception attack scenario simulated using \texttt{SQUANCH}. This protocol the same as Program \ref{prg:superdense_coding_demo} except that Eve intercepts Alice's transmitted qubits, measures them, and re-transmits them to Bob.}
	\label{prg:man_in_the_middle_demo}
\end{code}

\subsection{Quantum error correction}

\begin{figure}[h]
\scriptsize
\mbox{
\Qcircuit @C=1em @R=.7em {
&\push{\rule{1em}{0em}}&&\mbox{Alice}&&\push{\rule{0em}{1em}}&        \push{\rule{3em}{0em}} &\push{\rule{1em}{0em}}&&&\mbox{Bob}&&&&\push{\rule{0em}{1em}}\\
\lstick{|\psi\rangle} & \ctrl{3} & \ctrl{6} & \gate{H} & \ctrl{1} & \ctrl{2} & \multigate{8}{E} & \ctrl{1} & \ctrl{2} & \targ     & \gate{H} & \ctrl{3} & \ctrl{6} & \targ     & \rstick{\ket{\psi}} \qw\\
\lstick{\ket{0}}      & \qw      & \qw      & \qw      & \targ    & \qw      & \ghost{E}        & \targ    & \qw      & \ctrl{-1} & \qw      & \qw      & \qw      & \qw       & \qw \\
\lstick{\ket{0}}      & \qw      & \qw      & \qw      & \qw      & \targ    & \ghost{E}        & \qw      & \targ    & \ctrl{-2} & \qw      & \qw      & \qw      & \qw       & \qw \\
\lstick{\ket{0}}      & \targ    & \qw      & \gate{H} & \ctrl{1} & \ctrl{2} & \ghost{E}        & \ctrl{1} & \ctrl{2} & \targ     & \gate{H} & \targ    & \qw      & \ctrl{-3} & \qw \\
\lstick{\ket{0}}      & \qw      & \qw      & \qw      & \targ    & \qw      & \ghost{E}        & \targ    & \qw      & \ctrl{-1} & \qw      & \qw      & \qw      & \qw       & \qw \\
\lstick{\ket{0}}      & \qw      & \qw      & \qw      & \qw      & \targ    & \ghost{E}        & \qw      & \targ    & \ctrl{-2} & \qw      & \qw      & \qw      & \qw       & \qw \\
\lstick{\ket{0}}      & \qw      & \targ    & \gate{H} & \ctrl{1} & \ctrl{2} & \ghost{E}        & \ctrl{1} & \ctrl{2} & \targ     & \gate{H} & \qw      & \targ    & \ctrl{-6} & \qw \\
\lstick{\ket{0}}      & \qw      & \qw      & \qw      & \targ    & \qw      & \ghost{E}        & \targ    & \qw      & \ctrl{-1} & \qw      & \qw      & \qw      & \qw       & \qw \\
\lstick{\ket{0}}      & \qw      & \qw      & \qw      & \qw      & \targ    & \ghost{E}        & \qw      & \targ    & \ctrl{-2} & \qw      & \qw      & \qw      & \qw       & \push{\rule{0em}{1em}} \qw
\gategroup{1}{2}{10}{6}{1em}{.}
\gategroup{1}{8}{10}{15}{1em}{.}
}
}
\caption{Circuit diagram of encoding and decoding qubits using the Shor code. $E$ represents a quantum channel with an error model which can corrupt a single physical qubit by applying a random unitary operator.}
\label{fig:shor_code}
\end{figure}
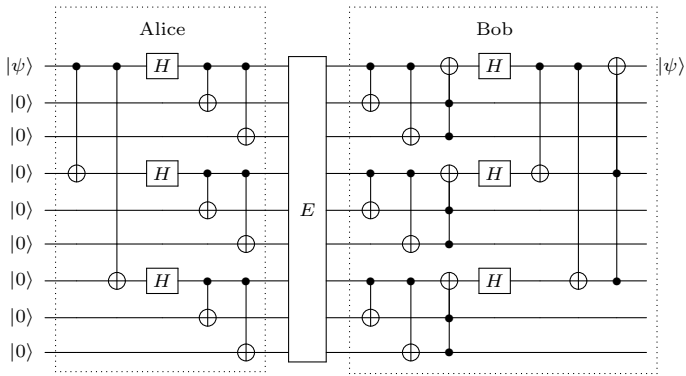

When qubits are transmitted over quantum channels, they are subject to a complex set of errors which can cause them to decohere, depolarize, or be lost to the environment. For quantum information transfer to be feasible, the information must be encoded in a error-resistant format using any of a variety of quantum error correction models. In this demonstration, we show how to use \texttt{SQUANCH}’s channel and error modules to simulate quantum errors in a transmitted message, which we correct for using the Shor Code, shown as a circuit diagram in Figure \ref{fig:shor_code}. This error correction model encodes a single logical qubit into the product of 9 physical qubits and is capable of correcting for arbitrary single-qubit errors.

In this demo, we have two pairs of agents: Alice and Bob will communicate a message which is error-protected using the Shor code, and DumbAlice and DumbBob will transmit the message without error correction. Formally, for each state $\ket{\psi}$ to be transmitted through the channel, the following procedure is simulated:

\begin{enumerate}
    \item Alice has some state $\ket{\psi}=\alpha_0\ket{0}+\alpha_1\ket{1}$, which she wants to send to Bob through a noisy quantum channel. She encodes her single-qubit state in nine logical qubit as $\ket{\psi} \mapsto \alpha_0\bigotimes_{j=1}^3\frac{1}{\sqrt{2}}\left(\ket{000}+\ket{111}\right) + \alpha_1\bigotimes_{k=1}^3\frac{1}{\sqrt{2}}\left(\ket{000}-\ket{111}\right)$ using the circuit diagram in Figure \ref{fig:superdense_coding_circuit}.
    \item DumbAlice wants to send the same state, but she doesn't error-protect the state and transmits the unencoded state $\ket{\psi}\otimes{\ket{00\cdots0}}$.
    \item Alice and DumbAlice send their qubits through the quantum channel $E$ to Bob and DumbBob, respectively. The channel may apply an arbitrary unitary operation to a single physical qubit in each group of nine.
    \item Bob receives Alice's qubits and decodes them using the Shor decoding circuit shown in Figure \ref{fig:shor_code}. 
    \item DumbBob expects $\ket{\psi}\otimes{\ket{00\cdots0}}$ from DumbAlice and only measures the results of the the first qubit in each group of nine.
\end{enumerate}

Transmitting an image is unsuitable for this scenario due to the larger size of the Hilbert space involved compared to the previous two demonstrations. (Each \texttt{QSystem.state} for $N=9$ uses 2097264 bytes, compared to 240 bytes for $N=2$.) Instead, Alice and DumbAlice will transmit the bitwise representation of a short message encoded as $\sigma_z$-eigenstates, and Bob and DumbBob will attempt to re-assemble the message. 

Since this demonstration is fairly long, we'll split the code into two parts. First, we define the Shor encoding and decoding circuits as ordinary Python functions:

\begin{code}
	\pythonfile{code/shor_code.py}
	\caption{Implementation of the nine-qubit Shor code in \texttt{SQUANCH}.}
	\label{prg:shor_code_demo}
\end{code}

Assuming that the functions defined in Program \ref{prg:shor_code_demo} are imported, we can define the agent logic and quantum channel model in Program \ref{prg:quantum_error_correction_demo}.

\begin{code}
	\pythonfile{code/quantum_error_correction.py}
	\caption{Demonstration of quantum error correction using the Shor code.}
	\label{prg:quantum_error_correction_demo}
\end{code}

\begin{figure}[h]
    \centering
    \includegraphics[width=\columnwidth]{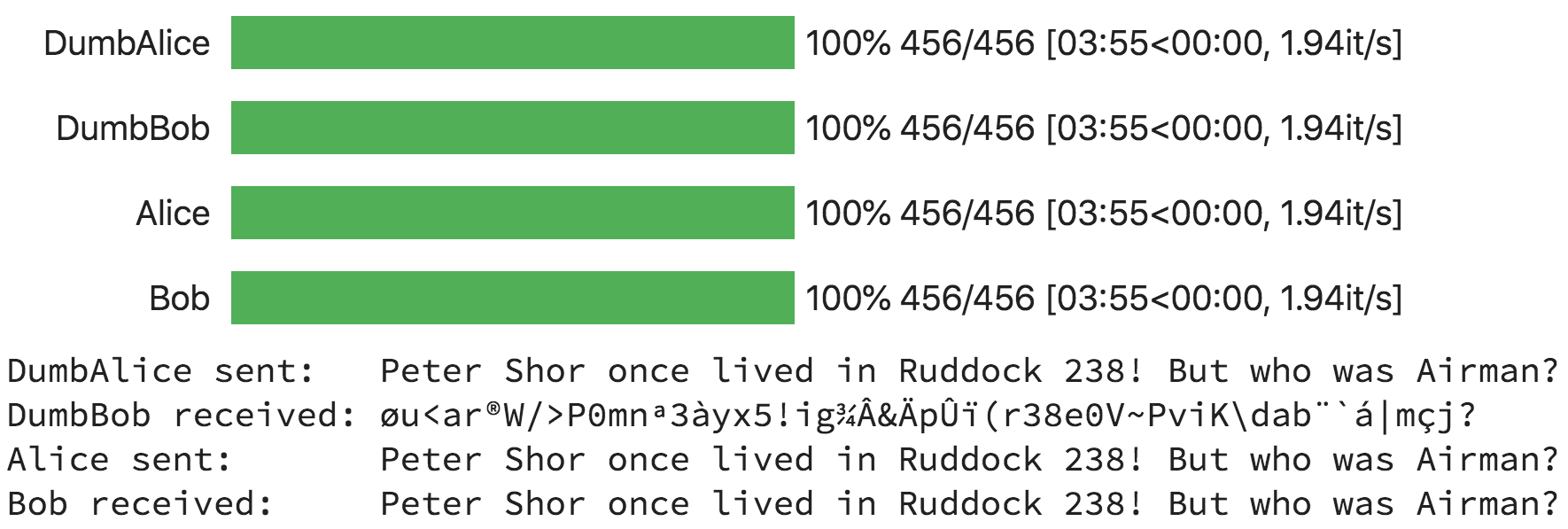}
    \caption{A screenshot of the output of Program \ref{prg:quantum_error_correction_demo} running in a Jupyter notebook. The original message can be recovered if a quantum error correcting protocol is used. Progress bars are automatically updated from \texttt{Agent.qstream.\_\_iter\_\_} when running programs with \texttt{Simulation(...).run()} and are shown in notebook or terminal environments.}
    \label{fig:quantum_error_correction_results}
\end{figure}

The results of Program \ref{prg:quantum_error_correction_demo} are shown in Figure \ref{fig:quantum_error_correction_results}. (A screenshot is provided as Unicode characters are problematic to include in \LaTeX documents.)

\section{Conclusions}

We have introduced \texttt{SQUANCH}, an open-source Python framework for creating distributed and performant simulations of multi-party quantum networks. The framework includes modules for representing quantum information at the single-particle, multi-particle, and ensemble levels, and large number of quantum gates for manipulating quantum information, enumerated in full in Appendix \ref{sec:gatesList}. The higher-abstraction modules in the framework provide software infrastructure for constructing the nodes of a quantum network by programming the actions of a quantum-mechanical ``agent'' and for connecting the network with noisy information channels. Virtually all components of the framework are configurable or extensible, allowing users to program simulations ranging from simple demonstrations to complex and detailed computational models. We hope that the development of this framework will stimulate exploration in the exciting field of quantum networking algorithms.

\section*{Acknowledgements}

We wish to thank our colleagues at INQNET, Rishiraj Pravahan, Yewon Gim, and Maria Spiropulu for providing support and helpful feedback throughout development. We also thank Sunil Pai for reviewing this whitepaper. All quantum circuits depicted in this paper were drawn using the \texttt{Q-circuit} \LaTeX package \cite{Eastin2004Q-circuitTutorial}.

\bibliographystyle{ieeetr}
\bibliography{main.bbl}

\appendix

\begin{table*}[h]
\setlength{\tabcolsep}{.5em} 
\begin{tabular}{lllc}
\toprule
Operation & Name & Definition & Symbol
\\\midrule
\texttt{H(qubit)} & Hadamard gate & $H=\frac{1}{\sqrt{2}}\smat{1}{1}{1}{-1}$ & \Qcircuit@C=1em@R=.7em{&\gate{H}&\qw} \\[1.5em]
\texttt{X(qubit)} & Pauli-X gate & $\sigma_x = \smat{0}{1}{1}{0}$ & \Qcircuit@C=1em@R=.7em{&\targ&\qw} \\[1.5em]
\texttt{Y(qubit)} & Pauli-Y gate & $\sigma_y = \smat{0}{-i}{i}{0}$ & \Qcircuit@C=1em@R=.7em{&\gate{Y}&\qw} \\[1.5em]
\texttt{Z(qubit)} & Pauli-Z gate & $\sigma_z = \smat{1}{0}{0}{-1}$ & \Qcircuit@C=1em@R=.7em{&\gate{Z}&\qw} \\[1.5em]
\texttt{RX(qubit, angle)} & Rotation-X gate & $R_x(\theta) = \cos\frac{\theta}{2}\id - i \sin\frac{\theta}{2}\sigma_x$ & \Qcircuit@C=1em@R=.7em{&\gate{R_x(\theta)}&\qw} \\[1.5em]
\texttt{RY(qubit, angle)} & Rotation-Y gate & $R_y(\theta) = \cos\frac{\theta}{2}\id - i \sin\frac{\theta}{2}\sigma_y$ & \Qcircuit@C=1em@R=.7em{&\gate{R_y(\theta)}&\qw} \\[1.5em]
\texttt{RZ(qubit, angle)} & Rotation-Z gate & $R_z(\theta) = \cos\frac{\theta}{2}\id - i \sin\frac{\theta}{2}\sigma_z$ & \Qcircuit@C=1em@R=.7em{&\gate{R_z(\theta)}&\qw} \\[1.5em]
\texttt{PHASE(qubit, angle)} & Phase shift gate & $R_\phi = \smat{1}{0}{0}{e^{i\phi}}$ & \Qcircuit@C=1em@R=.7em{&\gate{R_\phi}&\qw} \\[1.5em]
\texttt{CNOT(ctrl, targ)} & Controlled-NOT gate & $\mathrm{CNOT}_{jk} = \ketbra{0}{0}_j \otimes\id_k + \ketbra{1}{1}_j \otimes\sigma_{x;k}$ & \tab{\Qcircuit@C=1em@R=.7em{&\ctrl{1}&\qw\\&\targ&\qw}} \\[1.5em]
\texttt{CPHASE(ctrl, targ, angle)} & Controlled-phase gate & $\mathrm{CPHASE}_{jk}(\theta) = \ketbra{0}{0}_j \otimes\id_k + \ketbra{1}{1}_j \otimes\sigma_{x;k}$ & \tab{\Qcircuit@C=1em@R=.7em{&\ctrl{1}&\qw\\&\gate{R_\phi}&\qw}} \\[1.5em]
\texttt{CU(ctrl, targ, unitary)} & Controlled-unitary gate & $\mathrm{C}_j(U)_k = \left(\ketbra{0}{0}\right)_j \otimes\id_k + \left(\ketbra{1}{1}\right)_j \otimes U_{k}$ & \tab{\Qcircuit@C=1em@R=.7em{&\ctrl{1}&\qw\\&\gate{U}&\qw}} \\[1.5em]
\texttt{SWAP(qubit1, qubit2)} & SWAP gate & $\mathrm{SWAP}_{jk} = \mathrm{CNOT}_{kj} \mathrm{CNOT}_{jk} \mathrm{CNOT}_{kj}$ & \tab{\Qcircuit@C=1em@R=1.2em{&\qswap&\qw\\&\qswap \qwx&\qw}} \\[2em]
\texttt{TOFOLLI(ctrl1, ctrl2, targ)} & Toffoli gate & $\mathrm{CCNOT}_{ijk} = $\tab{\scriptsize\shortstack[l]{$\ketbra{0}{0}_i\ketbra{0}{0}_j \otimes\id_k + \ketbra{0}{0}_i\ketbra{1}{1}_j \otimes\id_k + $\\ $\ketbra{1}{1}_i \ketbra{0}{0}_j \otimes\id_k + \ketbra{1}{1}_i\ketbra{1}{1}_j \otimes\sigma_{x;k} $}} & \tab{\Qcircuit@C=1em@R=.7em{&\ctrl{1}&\qw\\&\ctrl{1}&\qw\\&\targ&\qw}} \\[2em]
\texttt{qubit.measure()} & Measurement & $\hat\rho \mapsto \hat\rho_i = \frac{\ketbra{i}{i} \hat\rho \ketbra{i}{i}^\dagger}{\tr [\hat\rho \ketbra{i}{i}]},\,$ return bit $i$ & \Qcircuit@C=1em@R=.7em{&\meter\qw&\cw} \\[1.5em]
\bottomrule
\end{tabular}
\caption{A list of the built-in quantum operations included in \texttt{SQUANCH} as of version 1.1.0. All gates take at least one \texttt{Qubit} argument and return nothing, modifying the state of the associated \texttt{QSystem} in-place.}
\label{table:gates}
\end{table*}

\section{Full source code}
\label{sec:sourceCode}

The full source code for the \texttt{SQUANCH} framework is available on GitHub at \url{github.com/att-innovate/squanch} or from the Python Package Index at \url{pypi.org/project/SQUANCH}. Documentation is available on the documentation website at \url{att-innovate.github.io/squanch}. The full source code for each of the demonstrations in Section \ref{sec:demos} is available in the \texttt{/demos} directory of the GitHub repository.

\section{Formal definition of a quantum agent}
\label{sec:quantum_agent_definition}

 Our definition for a quantum agent is chosen to mirror the structure of the \texttt{squanch.Agent} class, but is essentially a networked quantum Turing machine. A quantum agent is represented by a 5-tuple $\left(C, M_C, M_Q, G, T\right)$:
\begin{itemize}
    \item $C = (I_C, I_Q, O_C, O_Q)$ is a configuration of channels connecting the agent to other agents. $I_Q$ and $O_Q$ are input and output quantum channels, through which the agent can receive and send qubits, and $I_C$ and $O_C$ are input and output classical channels\footnote{Most definitions of quantum channels include the ability to transmit classical information; we make the explicit distinction between quantum and classical channels here for better consistency with the structure of the \texttt{squanch.Agent} class.}.
    \item $M_C$ is the classical memory, where the $k$th bit is indexed as $M_C [k]$.
    \item $M_Q$ is the quantum memory, which can store an arbitrary number of qubits, where the $k$th qubit is $M_Q [k]$. Qubits received through $I_Q$ can be stored in $M_Q$ until they are sent through $O_Q$, and the measurement result of a qubit can be stored in $M_C$.
    \item $G$ is an arbitrary but finite\footnote{Some gates, such as the \texttt{PHASE} gate, take parameters, which at first glance seems to make it problematic for the transition function $\delta$ to be able to access them from a finite set. However, assuming $G$ contains a functionally complete set of quantum gates, compositions of $g\in G$ can be used to approximate any unitary operator to within arbitrary accuracy \cite{Barenco1995ElementaryComputation}.} set of quantum operations. In addition, we always include ``operators'' for measurement and for sending qubits to a connected agent.
    \item A classical Turing machine\footnote{This is not technically a Turing machine from a computability standpoint, since $T$ has access to a random oracle by way of the measurement operation.} $T=(Q,\Sigma,\delta)$ where:
    \begin{itemize}
        \item $Q$ is a finite set of classical states
        \item $\Sigma$ is a finite tape alphabet with a blank symbol
        \item $\delta : Q \times \Sigma \rightarrow \Sigma  \times Q \times \{L,R\} \times \{L_Q,R_Q\} \times G$ is a modified transition function which also includes the traversal of quantum memory with $\{L_Q, R_Q\}$ and the ability to perform an operation $g\in G$ on the current qubit.
    \end{itemize}
\end{itemize}

In the non-networked case that $C=(\emptyset, \emptyset, \emptyset, \emptyset)$, the Agent has full access to the state space of their quantum system, and this model is seemingly reducible to the definition of a quantum Turing machine given in \cite{Bernstein1997QuantumTheory}, although a formal proof of this is beyond the scope of this paper.

\section{List of gates}
\label{sec:gatesList}

A complete list of gates available in \texttt{SQUANCH v1.1.0} is given in Table \ref{table:gates}.

\section{\texttt{SQUANCH} API reference}
\label{sec:api}

An abridged listing of the \texttt{SQUANCH} API is given on the remaining pages of this manuscript. Refer for the documentation for a more complete version.

\newpage 

\foreach \x in {1,...,4}
{
\begin{figure*}
    \centering
    \includegraphics[page=\x,width=\textwidth, trim=2.5cm 2.5cm 2.5cm 2.5cm]{SQUANCH.pdf}
\end{figure*}
}

\end{document}